\documentclass[12pt]{article}
\usepackage{a4wide}

\usepackage{amsmath}
\usepackage{amssymb}
\usepackage[T1]{fontenc}
\usepackage{color}

\newcommand{\pr}{{}^{\prime}}
\newcommand{\dpr}{{}^{\prime\prime}}
\newcommand{\mg}{m_{\tilde{G}}}
\newcommand{\tg}{\tau_{\tilde{G}}}

\begin{document}

\begin{center}
{\large {\bf 3.5 keV x-ray line from decaying gravitino dark matter}}\\
\vspace*{1cm}
{\bf N.-E.\ Bomark\footnote{nilserik.bomark@gmail.com} and L.\
  Roszkowski\footnote{Leszek.Roszkowski@fuw.edu.pl. On leave of
    absence from the University of Sheffield, UK.}} \\
\vspace{0.3cm}
National Centre for Nuclear Research, Ho\.za 69, 00-681 Warsaw, Poland
\end{center}

\begin{abstract}
  Extremely weakly interacting particles like the gravitino may be stable
  enough on cosmological time scales to constitute a good dark matter candidate
  even in the presence of $R$-parity violation.  We consider the
  possibility that the recently identified 3.5~keV x-ray line can be
  generated in light gravitino decays to neutrinos and photons. We find
  that this is indeed possible in loop processes induced by trilinear
  lepton-number-violating couplings. We show that in order to avoid overproduction
  of gravitinos, the reheating temperature has to be at most around 100 GeV to 1 TeV. 
  Finally we briefly discuss associated LHC
  phenomenology due to a relatively light gluino and
  multijet/multilepton events from $R$-parity violating decays of neutralinos.

\end{abstract}

\section{Introduction}
Despite the abundance of gravitational evidence of the existence of
dark matter (DM), a  confirmed detection signal through nongravitational
modes is still lacking. However, with the improving sensitivity in cosmic-ray
measurements, direct detection experiments and increasing reach in
collider searches, a genuine dark matter signal could be expected.

In fact, recent studies of stacked x-ray spectra from the XMM-Newton
telescope, have revealed an unidentified line with the central energy
of $3.5$~keV~\cite{Bulbul,Boyarsky}. However, one should bear in mind
that the significance of the signal is not that high yet ($\simeq
$4-5$\ \sigma$~\cite{Bulbul}) and that, although currently lacking, more conventional explanations of the
line in terms of atomic physics effects have not been ruled out. On the other hand, it is tempting
to consider more exotic explanations of the signal in terms of
decaying or annihilating dark matter since a monochromatic photon line
signal would be a smoking gun of dark matter. While annihilating dark
matter does not seem compatible  with the
signal~\cite{Sanninoetal}, it is possible to explain it with eXciting
dark matter~\cite{XDM}, where the photons come from the transition
from the excited state down to the ground state for the dark matter
particle, which in this case can be significantly heavier than 3.5~keV.

Interpretations in terms of light decaying dark matter seem more
promising for meeting the conditions implied by the data. The required properties of such dark matter are~\cite{Bulbul,Boyarsky}

\begin{eqnarray}
  m_{\text{DM}}    & \simeq & 7 \text{ keV}, \\\nonumber
  \tau_{\text{DM}} & \simeq &  10^{28}\text{ s}.
\end{eqnarray}

In fact, several candidates have already been
considered in this context, and  sterile
neutrinos~\cite{SterNeu}, axions~\cite{Axion} and axinos~\cite{Axino}
have already been demonstrated to be able to produce the observed line
through their decay. Also decaying moduli~\cite{Moduli} and
millicharged dark matter~\cite{Millicharge} as well as multicomponent
dark matter~\cite{MultiCompDM} have been shown to be compatible with the
data.

In this paper, we consider the possibility that the x-ray line is
produced by the gravitino decaying through $R$-parity-violating (RPV)
processes to, e.g., a photon and a neutrino.
While LHC data imply that the masses of ordinary sparticles are in the
TeV range, this does not necessarily apply to the gravitino since its
mass is set by the scale of supersymmetry breaking and it does not
have to be similar to the other sparticles.  As a matter of fact, in
gauge-mediated supersymmetry breaking a keV-scale gravitino mass is
natural~\cite{GMSB}.

It is also worth mentioning that such a light gravitino would
potentially constitute warm dark matter~\cite{GravWDM} and hence help
solve some possible problems of the cold dark matter paradigm, most
notably the cusp-core problem and the missing satellite problem,
although other solutions are also possible.

The rest of this paper is structured as follows. In
Section~\ref{sec:Decay} we discuss the decay channels for a light
gravitino in $R$-parity-violating supersymmetric models and calculate
the relevant decay widths. In Section~\ref{sec:LHCrelic} we discuss
the compatibility of this scenario with early Universe cosmology,
especially the issue of not over-closing the Universe with gravitinos,
and we make comments about associated LHC phenomenology. Section~\ref{sec:Conclusion} contains our conclusions.

\section{$R$-parity-violating decay of gravitino dark matter}\label{sec:Decay}
If the requirement of $R$-parity conservation is lifted, a number of
terms become allowed in the superpotential. These are trilinear and
bilinear lepton-number-violating terms \cite{barb},
\begin{equation}\label{eq:Lviol}
    \lambda_{ijk} L_iL_j\bar{E}_k+\lambda\pr_{ijk}L_iQ_j\bar{D}_k+\mu_iH_1L_i,
\end{equation}
where $i,j,k=1,2,3$, $L_i,Q_i,H_1$ are left-chiral lepton, quark and
Higgs superfields and $E_i,D_i$ are right-chiral lepton and down-quark
superfields, as well as trilinear baryon-number-violating
terms
\begin{equation}\label{eq:Bviol}
    \lambda\dpr_{ijk}\bar{U}_i\bar{D}_j\bar{D}_k,
\end{equation}
where $D_i,U_i$ are right-chiral down- and up-quark superfields.

If $R$-parity is violated, supersymmetry does not provide any stable
dark matter candidate. However, the gravitino can, due to the
smallness of its interactions, still be long-lived enough to
constitute the missing matter of the Universe~\cite{TY,CM,BM}.

The question we want to address is  whether decaying gravitino dark
matter could be the cause of the 3.5 keV x-ray line recently
reported. The baryon-number-violating terms of Eq.~(\ref{eq:Bviol}) do
not induce any gravitino decay including a monochromatic photon line,
and hence we can ignore them from now on; note also that proton
stability requires the absence of combinations of baryon-number- and lepton-number-violating terms and therefore we shall assume the
terms of Eq.~(\ref{eq:Bviol}) to all be zero.

If $R$-parity is violated by the bilinear terms $\mu_i$, a decay
$\tilde{G}\rightarrow\nu \gamma$ is allowed which gives the required
signature for a gravitino mass of 7 keV. However, it turns out that
the lifetime associated with this decay is given by~\cite{TY}
\begin{equation}\label{eq:BiLintau}
\tg\approx 4\times 10^{11}\ \text{s\ } |U_{\nu\tilde\gamma}|^{-2}\left(\frac{\mg}{10 \text{ GeV}}\right)^{-3},
\end{equation}
where $U_{\nu\tilde\gamma}$ is the mixing between neutrinos and
photinos induced by the $\mu_i$ couplings and $\mg$ is the gravitino
mass. For the required gravitino mass of 7 keV,
Eq.~(\ref{eq:BiLintau}) gives $\tg\approx 1.1\times 10^{30}\ \text{s\
} |U_{\nu\tilde\gamma}|^{-2}$ which exceeds the values required to
explain the observed line. This was also pointed out
in Ref.~\cite{RPVdecays} where it was concluded that although there are
several sparticles capable of explaining the line signal in $R$-parity-violating supersymmetry, the gravitino is not one of them; however, as
we will see below, this conclusion changes if we also take trilinear
$R$-parity violation into account.

It was demonstrated in Ref.~\cite{LOR} that trilinear lepton-number-violating couplings $\lambda_{ijk}$ and $\lambda\pr_{ijk}$
allow for sfermion-fermion loops that also can produce the required
$\tilde{G}\rightarrow\nu \gamma$ decay.\footnote{At the same time this paper was published, it was demonstrated in Ref.~\cite{Axinoloop} that similar loops for decaying axinos are also compatible with the data.} In order for the loop decay to
be possible one needs two identical flavors in the operator so this
can only happen if $i=k$ or $j=k$ for $\lambda_{ijk}$ couplings, in
which case the loop will contain lepton and slepton lines of flavor
$k$, or if $j=k$ for $\lambda\pr_{ijk}$ couplings, in which case the
loop will consist of down-quark and squark propagators of flavor $k$.

The decay width due to the loop decays is given by~\cite{LOR}
\begin{equation}\label{eq:TriWidth}
  \Gamma_{\tilde G}=\frac{\alpha\lambda^2\mg}{2048\pi^4}\frac{m_f^2}{M_p^2}\overline{|\mathcal{F}|^2},
\end{equation}
where $M_p=2.4\times 10^{18}$ GeV is the reduced Planck mass, $m_f$ is
the mass of the fermion in the loop and $\overline{|\mathcal{F}|^2}$
is a loop factor given (with a minor correction as compared
to Ref.~\cite{LOR}\footnote{It is worth mentioning that in this type of loop calculations  a delicate issue is the relative sign between two sets of diagrams differing by the reversal of the charge flow in the loop. It was briefly stated in Ref.~\cite{LOR} that the end result is that the amplitudes of both diagrams add up constructively, giving a factor of 2 when the sfermions exchanged in the loops are degenerate in mass. To understand why these two sets add constructively, it is convenient to employ the formalism of Ref.~\cite{DEHK} to deal with the clashing arrows that appear in one of the sets of diagrams. One then obtains one relative minus sign from the photon coupling due to the reversal of the electric charge flow and, in addition, another relative minus sign coming from the reversal of the fermion propagator in the gravitino vertex, as can be seen by examining gravitino Feynman rules, e.g.\ in Appendix A.3.2 of Ref.~\cite{Boltz}.}) in Appendix~A of Ref.~\cite{BLOR}.

While the tree-level three-body decay through the trilinear terms give
a decay width $\propto \mg^{7}$, and hence becomes negligible for small
gravitino masses, the loop decay gives a width $\propto \mg$ and
therefore dominates at small masses~\cite{BLOR}. Note that this means
the loop decay also decreases much more slowly than the bilinear
induced decay when the gravitino mass is decreased, and therefore it opens
the possibility of explaining the observed excess.

For small gravitino masses the decay width is essentially independent
of the other sparticles' masses, the only dependence is on the RPV
coupling and the mass of the fermion in the loop. The latter
dependence is a consequence of the helicity structure of the diagram
that requires a helicity flip whose probability depends on the mass
term for the fermion. To calculate the lifetime of the gravitino we
use Eq.~(\ref{eq:TriWidth}) together with LoopTools~\cite{LoopTools}
for the loop factor $\overline{|\mathcal{F}|^2}$. The result is given
in Table~\ref{Tab:ReqCoupling} where the coupling strength required to
obtain a lifetime of $10^{28}\ $s for each of the particles that can
appear in the loop is shown.

\begin{table}[ht!dp]
\caption{The coupling strength required to obtain a lifetime of $10^{28}\ $s for the various particles that can appear in the loop. The third column gives the couplings that can give rise to the mentioned loops.}
\begin{center}
\begin{tabular}{|l|l|c|}
  \hline
  Particle in loop   &  $\lambda$ required for $\tg = 10^{28}\ $s  &  Couplings \\\hline
  electron & 23 & $\lambda_{121}$, $\lambda_{131}$\\
  muon & 0.11 & $\lambda_{122}$, $\lambda_{232}$\\
  tauon & 0.0066 & $\lambda_{133}$, $\lambda_{233}$\\
  $d$ quark & 1.9 & $\lambda\pr_{i11}$\\
  $s$ quark & 0.065 & $\lambda\pr_{i22}$\\
  $b$ quark & 0.0016 & $\lambda\pr_{i33}$\\
  \hline
\end{tabular}
\end{center}
\label{Tab:ReqCoupling}
\end{table}%

Since couplings larger than unity are problematic for perturbativity
reasons as well as in conflict with experiment for most couplings, we
see in Table~\ref{Tab:ReqCoupling} that electron or $d$ quark couplings
are too large.  However, all the other cases seem compatible with the
observed line. As expected the $b$ quark loop requires the smallest
coupling of only $2 \times 10^{-3}$. When comparing the values in
Table~\ref{Tab:ReqCoupling} with experimental
constraints~\cite{RPVConstr}, we see that at least the muon loop
contribution is in conflict with constraints for sparticle masses of
100 GeV; however, in light of the lack of detection at the LHC, such
low sparticle masses are not realistic and for sparticle masses above
1 TeV there is no conflict with the above-mentioned constraints.

There are also constraints coming from the neutrino masses~\cite{RPVneutrinoConstr}, but they are highly dependent on the full sparticle mass spectrum so there should always be room to meet those constraints.

From a model-building point of view, one often expects the flavor
structure of the trilinear $R$-parity-violating couplings to resemble
the flavor structure of the standard model Yukawa couplings and hence
one would expect that the couplings including heavy flavors
(especially $\lambda\pr_{i33}$) are the
largest~\cite{LolaModels}. This is in good agreement with our finding
that couplings with heavy flavors are the most suited to explain the 3.5~keV
x-ray line discussed here.

\section{LHC signatures and relic density of gravitinos}\label{sec:LHCrelic}
One important issue for gravitino dark matter is the production of the
correct relic density. The thermal relic density of gravitinos from scattering processes in the primordial plasma,
$\Omega_{\tilde G} h^2$, is given by the gravitino mass, $\mg$, the
reheating temperature, $T_R$, and the gluino mass, $m_{\tilde g}$,
according to~\cite{GravTR}
\begin{equation}\label{eq:ReheatTemp}
  \Omega_{\tilde G} h^2=0.27 \left(\frac{100 \text{ GeV}}{\mg}\right)\left(\frac{T_R}{10^{10}\text{ GeV}}\right)\left(\frac{m_{\tilde g}}{1 \text{ TeV}}\right)^2.
\end{equation}

From Eq.~(\ref{eq:ReheatTemp}) we see that for a gravitino as light as
7 keV, in order to avoid overproduction, one would prefer a light
gluino, however, a very light gluino would be incompatible with LHC constraints.

Without detailed knowledge of  mass hierarchies among the
sparticles, it is impossible to give an exact number for the limit on
the gluino mass. However, as a crude estimate we can assume that the
gluino is the lightest ordinary sparticle and it is being produced in pairs
that subsequently decay through a $\lambda\pr_{i33}$ coupling to
two $b$ jets and a neutrino. (Here we assume that the decay to a charged
lepton, a top-quark and a $b$ quark is essentially absent due to
phase-space suppression by the top mass, if this channel is
significant, it should if anything strengthen the constraint due to
its easier detection. For studies of this channel see Ref.~\cite{Gluinobtl}.) The resulting topology of four $b$ jets and missing transverse energy
has been searched for in the context of gluinos decaying to $b$ quark
pairs and neutralinos~\cite{ATLASbbchi,CMSgluino}. There are also
searches for gluinos decaying to quark pairs and
neutralinos~\cite{ATLASqqchi,CMSqqchi} and they are slightly more
constraining with the best limit (for zero neutralino mass) being
$m_{\tilde g}\gtrsim1400$ GeV.

Naively using this limit in Eq.(\ref{eq:ReheatTemp}) translates to $T_R \lesssim 170$ GeV; however, with $T_R < m_{\tilde g}$, the production of gravitinos would be exponentially (Boltzmann) suppressed as compared to Eq.(\ref{eq:ReheatTemp}), allowing for a significantly higher reheating temperature. How high this temperature can be will be very sensitive to the precise mass spectrum of all sparticles potentially capable of producing gravitinos through thermal scattering or decay and a detailed discussion of this goes beyond the scope of this paper. For the current discussion it suffices that there are reheating temperatures in the range 100 GeV to 1 TeV that, given an appropriate sparticle spectrum, can account for the right gravitino abundance~\cite{GravProd}.

It is also possible to have late-time entropy
creation from the decay of other relics~\cite{LateEntropy} which can
wash out the gravitino density and hence alleviate the constraint on
$T_R$ stated above.

In this scenario thermal leptogenesis is clearly impossible,
firstly because the reheating temperature is too low and while it is
possible to achieve a high enough reheating temperature by late-time
entropy production~\cite{GMSBLeptogen}, the lepton-number-violating
couplings will erase the lepton asymmetry before the electroweak phase
transition~\cite{LeptogenConstr}. However, the reheating temperature
potentially remains above the electroweak phase transition
temperature, possibly leaving some room for electroweak baryogenesis
to account for the baryon asymmetry of the Universe~\cite{EWBaryogen}.
Another viable option would be the Affleck-Dine mechanism~\cite{AffleckDine}.

The violation of $R$-parity has rather significant implications for LHC
phenomenology~\cite{DreinerRoss}. Most importantly the neutralino,
even if lighter than all other sparticles of relevance to the LHC, is
no longer stable and hence will not give rise to missing transverse
energy but will rather decay into final states of standard model
particles. For LHC studies of all possible decay topologies,
see Ref.~\cite{BDLO}. If the gravitino decay is due to a tau loop, the LHC
phenomenology of decaying neutralinos is promising; the final state
will consist of multiple leptons and missing transverse energy, and
although many of the leptons will be taus this would be clearly
detectable if the neutralino production (either direct or through
decay) is large enough. Even more promising would be the muon case
where a large part of the final-state leptons would be muons.

For squark loops the expected signature would be multiple jets and
some missing transverse energy (from neutrinos) and/or charged
leptons, which, although not as clean as the multilepton case, should
be clearly seen. The least promising option, which might also be the
most likely, would be a $b$ quark loop which leads to a final state of
$b$ jets and missing transverse energy. However, as seen above, such
final states are also within reach of the LHC experiments. Also note
that if the neutralino is heavy enough, it can decay to a charged
lepton, a top quark and a $b$ quark, and that should be a rather clean
signal even if the branching fraction is small due to phase-space
suppression.

\section{Conclusions}\label{sec:Conclusion}
The possible existence of a x-ray line consistent with decaying dark
matter is an intriguing possibility. In this paper we have demonstrated
that such a signal can indeed be interpreted in terms of decay products of
gravitino dark matter if $R$-parity is violated by trilinear lepton-number-violating couplings which  induce loop decays of the gravitinos to
neutrinos and photons. Couplings as small as $2\times 10^{-3}$ are
shown to be capable of explaining the observed line, which means all
present constraints are satisfied.

The low gravitino mass of 7 keV required to fit the line means that
the reheating temperature of the Universe cannot be too high; without
any late-time entropy production, we conclude that it can be at most 100 GeV to 1 TeV in order to avoid over-closure of the universe
by the gravitinos. This leaves no room for thermal leptogenesis to
account for the baryon asymmetry of the Universe (even if late-time
entropy production is included to raise the reheating temperature, the
$R$-parity-violating couplings will still erase any lepton asymmetry),
but electroweak baryogenesis or Affleck-Dine baryogenesis might still be viable.

One would also expect the gluino, or some other sparticles participating in the gravitino production, to not be too heavy and that,
together with the multilepton or multijet final states expected from
neutralino (or possibly gluino) decay through the $R$-parity-violating
couplings responsible for the gravitino decay, would be very promising
for future searches for supersymmetry signals with $R$-parity breaking at the LHC.

\vspace*{0.2 cm}
{\bf Acknowledgements.}
We would like to thank Smaragda Lola, Per Osland and Are Raklev for useful clarifications regarding the decay width of the gravitino.
This work has been funded in part by the Welcome Programme of the
Foundation for Polish Science. L.R. is also supported in part by the STFC
consortium grant of Lancaster, Manchester, and Sheffield Universities,
and by the EC 6th Framework Programme No. MRTN-CT-2006-035505.


\end{document}